\documentclass[10pt]{bmc_article}    

\usepackage{cite} % Make references as [1-4], not [1,2,3,4]
\usepackage{url}  % Formatting web addresses  
\usepackage{ifthen}  % Conditional 
\usepackage{multicol}   %Columns
\usepackage[utf8]{inputenc} %unicode support
\usepackage{alltt}
\newif\ifarxiv
\arxivtrue
\ifarxiv
\usepackage{graphicx}
\else

\def\includegraphics{}
\fi

\newboolean{publ}

\newenvironment{bmcformat}{\begin{raggedright}\baselineskip20pt\sloppy\setboolean{publ}{false}}{\end{raggedright}\baselineskip20pt\sloppy}

\begin{document}
\begin{bmcformat}


\title{biobambam: tools for read pair collation based algorithms on BAM files}

%%%%%%%%%%%%%%%%%%%%%%%%%%%%%%%%%%%%%%%%%%%%%%
%%                                          %%
%% Enter the authors here                   %%
%%                                          %%
%% Ensure \and is entered between all but   %%
%% the last two authors. This will be       %%
%% replaced by a comma in the final article %%
%%                                          %%
%% Ensure there are no trailing spaces at   %% 
%% the ends of the lines                    %%     	
%%                                          %%
%%%%%%%%%%%%%%%%%%%%%%%%%%%%%%%%%%%%%%%%%%%%%%

\author{German Tischler\correspondingauthor$^{1}$%
       \email{German Tischler\correspondingauthor - german.tischler@sanger.ac.uk}%
      \and
         Steven Leonard$^{1}$%
      }

%%%%%%%%%%%%%%%%%%%%%%%%%%%%%%%%%%%%%%%%%%%%%%
%%                                          %%
%% Enter the authors' addresses here        %%
%%                                          %%
%%%%%%%%%%%%%%%%%%%%%%%%%%%%%%%%%%%%%%%%%%%%%%

\address{%
    \iid(1)The Wellcome Trust Sanger Insitute, Wellcome Trust Genome Campus, Hinxton, Cambridge, UK
}%

\maketitle

%%%%%%%%%%%%%%%%%%%%%%%%%%%%%%%%%%%%%%%%%%%%%%
%%                                          %%
%% The Abstract begins here                 %%
%%                                          %%
%% The Section headings here are those for  %%
%% a Research article submitted to a        %%
%% BMC-Series journal.                      %%  
%%                                          %%
%% If your article is not of this type,     %%
%% then refer to the Instructions for       %%
%% authors on http://www.biomedcentral.com  %%
%% and change the section headings          %%
%% accordingly.                             %%   
%%                                          %%
%%%%%%%%%%%%%%%%%%%%%%%%%%%%%%%%%%%%%%%%%%%%%%

\begin{abstract}
        % Do not use inserted blank lines (ie \\) until main body of text.
        \paragraph*{Background:} Sequence alignment data is often
        ordered by coordinate (id of the reference sequence plus position on the
        sequence where the fragment was mapped) when stored in BAM files, as
        this simplifies the extraction of variants between the mapped data
        and the reference or of variants within the mapped data. 
        In this order paired reads are usually separated in
        the file, which complicates some other applications 
        like duplicate marking or conversion to the FastQ format
	which require to access the full information of the pairs.
      
        \paragraph*{Results:} In this paper we introduce biobambam, an API
	for efficient BAM file reading supporting the efficient collation of
	alignments by read name without performing a complete resorting of
	the input file and some tools based on this API performing 
	tasks like marking duplicate reads and conversion to the FastQ
	format.

        \paragraph*{Conclusions:} In comparison with previous approaches
        to problems involving the collation of alignments by read name like the 
        BAM to FastQ or duplication marking utilities in the Picard suite 
        the approach of biobambam can often perform an equivalent task 
        more efficiently in terms of the required main memory and run-time.

\end{abstract}

\ifthenelse{\boolean{publ}}{\begin{multicols}{2}}{}

%%%%%%%%%%%%%%%%%%%%%%%%%%%%%%%%%%%%%%%%%%%%%%
%%                                          %%
%% The Main Body begins here                %%
%%                                          %%
%% The Section headings here are those for  %%
%% a Research article submitted to a        %%
%% BMC-Series journal.                      %%  
%%                                          %%
%% If your article is not of this type,     %%
%% then refer to the instructions for       %%
%% authors on:                              %%
%% http://www.biomedcentral.com/info/authors%%
%% and change the section headings          %%
%% accordingly.                             %% 
%%                                          %%
%% See the Results and Discussion section   %%
%% for details on how to create sub-sections%%
%%                                          %%
%% use \cite{...} to cite references        %%
%%  \cite{koon} and                         %%
%%  \cite{oreg,khar,zvai,xjon,schn,pond}    %%
%%  \nocite{smith,marg,hunn,advi,koha,mouse}%%
%%                                          %%
%%%%%%%%%%%%%%%%%%%%%%%%%%%%%%%%%%%%%%%%%%%%%%

%%%%%%%%%%%%%%%%
%% Background %%
%%
\section*{Background}

  The SAM (Sequence Alignment/Matching) and BAM (Binary Alignment/Matching) file formats
  have become the standard formats for storing sequence data which was obtained
  through high throughput sequencing and alignment of the resulting data to a reference genome.
  Both formats were introduced as part of the SAMtools package
  (cf.~\cite{Li15082009}). SAM is a human readable text format whereas BAM
  is a more compact and compressed binary format. The current specifation of the formats is
  availble at \cite{samfilespec}. These files can be used for many
  applications including the detection of variants between the contained
  data and a reference, sequencing quality control and long term storage.
  Many programs have been created for the alignment of sequencing reads to
  reference sequences including SSAHA \cite{ssaha}, BWA
  \cite{bwa,bwasw}, Bowtie \cite{bowtie,bowtie2}, SOAP \cite{soap,soap2}
  and SMALT \cite{smalt} and most of the recently published aligners are
  either capable of generating SAM or BAM output or come with a script for
  converting their output to SAM or BAM. Most sequence data produced at the
  time being is sequenced as paired end reads. Short linear DNA templates
  are sequenced from both ends. This produces a pair of reads for each
  template. Both ends of the pair are assigned the same read name in the resulting data
  files thus providing the information that both ends are most likely within a certain
  expected distance in the underlying genome. This information aids in
  correctly aligning the resulting short sequences to a reference or
  assembling the fragments to a new draft reference. In the data obtained
  from a sequencer the pairs are usually collated in some form, either the
  two ends of a pair directly follow each other in a file or appear in an
  equivalent position in two separate files such that each of the two holds
  only the information for one of the two ends. The order of reads aligned
  to a reference which is most suitable for calling variants between the
  reads and the reference or within the reads is however the one resulting
  from sorting the data by coordinate (id of the reference sequence plus position on the
  sequence where the fragment was mapped). Thus many SAM and BAM files
  are processed in this order. There are however some applications which
  require the complete information from each pair. This includes the
  conversion of BAM files to a FastQ format suitable for realignment or a
  de novo assembly for an alternative detection of variants (see e.g.~\cite{pmid22569178})
  as well as the marking of duplicate reads. It is thus useful to have
  a quick, easy to use and reliant way of collating reads from a SAM/BAM file
  by their name without needing to resort to a full resorting of the file by
  read names. For the application of duplicate marking it is in addition
  desirable to keep the order after collation as close to the coordinate
  sorted order as possible, as clusters of reads pairs mapped to the same
  coordinates need to be detected. In this paper we present
  \texttt{biobambam}, a C++ API for efficient read name collation in BAM file
  and two tools \texttt{bamtofastq} and \texttt{bammarkduplicates} based on
  this API. These tools are more efficient in terms of runtime and memory
  usage then previous tools solving equivalent tasks like the SamToFastq and MarkDuplicates
  modules in Picard (see \cite{picard}).
  \pb
  %
  %Finally we will present the two programs \texttt{bamtofastq} and \texttt{bammarkduplicates} using the
  %collation code and compare their performance in terms of memory usage and
  %speed to equivalent solutions achieved by other widely used tool sets like
  %Picard (see \cite{picard}).\pb
 
%%%%%%%%%%%%%%%%%%%%%%%%%%%%
%% Results and Discussion %%
%%
\section*{Implementation}

  The biobambam package is split into two parts. The front-end tools
  \texttt{bamtofastq} and \texttt{bammarkduplicates} show-casing some 
  applications of fast collation of alignments by name can be  found in the biobambam 
  source package (cf.~\cite{biobambam}). The implementation of the
  collation code and the BAM file input and output routines are part of the
  larger \texttt{libmaus} project (see \cite{libmaus}), which also contains some
  supporting data structures.\pb

  There are various code bases and APIs available for
  SAM and BAM file input and output, including SAMtools (C), SeqAn (C++, cf.~\cite{seqan})
  and Bio-samtools (Ruby, cf.~\cite{22640879}).
  We use our own implementation for reading BAM files, which can be found in the libmaus
  project (C++, \cite{libmaus}). The libmaus project also contains various
  supporting data structures which we use, including the biobambam API in its
  namespace \texttt{libmaus::bambam}. The front-end programs can be found in
  the biobambam project. The tools can easily be extended to handle the newer
  CRAM format (cf.~\cite{HsiYangFritz01052011}) via the \texttt{io\_lib} part
  of the Staden package (cf.~\cite{iolib,staden}) which contains the Gap5
  software (see \cite{gap5}).
  So far this extension has been implemented for the \texttt{bamtofastq} 
  program, which is capable of transforming CRAM files to FastQ.\pb
  
  In the following we will first describe the algorithms and data structures
  used to for collating alignments by their name. Subsequently we will
  present the API making the functionality available to other users. 

  \subsection*{Algorithms and Data Structures for Collation by Read Name}

  Although the BAM file format can store alignments in any order, most BAM
  files will either have the alignments collated by the corresponding read
  names or will contain the alignments sorted by their coordinates on the
  reference the reads were aligned to. 
  The first case will commonly appear as the output of alignment programs
  or if raw FastQ files coming from a sequencer are converted to the BAM format 
  without aligning the contained reads to any reference.
  In this setting the output of the alignments in an order collated by read name 
  to another format offering the same or less information is very simple.
  In the second case a straight-forward but often inefficient way is to
  first sort the input BAM file by query name using tools like SAMtools or
  Picard and then resort to a conversion as employed in the first case, as a BAM
  file sorted by query name will have the alignments collated by read name.
  For a BAM file sorted by alignment coordinates collating the alignments 
  by read name can often be done more efficiently by observing that while the
  alignments paired by read name will commonly not be consecutive in the file,
  they are in most cases close together. 
  If we denote the average coverage of a coordinate on the reference by $d$ 
  (i.e.~each coordinate is covered by $d$ reads/alignments on average), the 
  average absolute template length of a pair with both ends mapped to the
  same sequence by $t$ (e.g.~the absolute value of the distance between 
  the mapping positions of the $5'$ ends for Illumina paired-end reads) 
  and the read length by $l$, then we would expect the distance between 
  two such ends in the BAM file to be $\frac{d}{l} ( t - l )$ on average.
  The mean number of read ends starting at each position on the
  forward strand is $\frac{d}{l}$ and the distance between the two starting
  points on the forward strand is $t-l$. 
  Mapping the data from the whole human genome sequencing study ERP001231
  (cf.~\cite{ERP001231}) to the human genome (GRCh37, see \cite{GRCh37})
  using the SMALT aligner (see \cite{smalt}) for instance, we observe an average
  sequencing depth of $d=45$ with an average template length of $t=324$ at
  a mean read length of $l=101$ ($100$ base pairs were sequenced at one end of
  the templates and $102$ from the other end). 
  According to our formula this implies an average number of about $99$ 
  alignments between the two alignments of one pair in a BAM file containing the 
  aligned reads. The actual median we observe in the file is $107$. 
  Due to some improperly mapped pairs in the file the weighted average value we
  see is not a meaningful number.
  Thus for the average case it would be sufficient to use any type of data
  structure which allows fast insertion, deletion and lookup of alignments
  by read name for a small set of alignments.\pb
  One such data structure would be a hash table with collision resolution by
  separate chaining.
  In practice however we see cases where some read ends stay in this hash table
  for an extended time when we process a BAM file sorted by coordinates from
  start to end.
  This may happen for reads where the two ends map to different chromosomes (split reads).
  %or if against the best practice for BAM unmapped mates of a mapped 
  %end are not kept right next to their mate.
  %
  There are also often regions in a genome where the sequencing depth is
  much higher than on average, which can lead to a drastic increase in the
  amount of memory required to store the hash table at certain points.
  Instead of using a hash table with collision resolution we use a hash table
  $H$ of fixed size $h$ without collision resolution.
  If there is a collision because two alignments with different names are assigned 
  the same hash value, then the alignment previously in the hash table is removed
  from the table and inserted into a list $L$ of fixed size $l$ of alignments to
  be handled later.
  Each time the list $L$ runs out of space we sort the alignments in $L$ by
  read name. 
  This sorting may yield some new pairs, which we extract before storing the 
  unpaired alignments still in $L$ in a temporary file and emptying the list $L$.
  As soon as all alignments have been read from the source BAM file we move all the alignments
  remaining in $H$ over to the list $L$ and in the end flush the list $L$ by
  sorting the remaining elements by name, extracting the resulting pairs and 
  writing the remaining unpaired alignments to another temporary file. 
  As all the temporary files are sorted by name, we can
  easily merge the files together to obtain a stream of alignments that is
  sorted by read name. In this stream it is again simple to detect and
  output pairs. A diagram of this data flow can be seen in Figure
  1. % \ref{dsfig}.
  Using this kind of setup we are able to quickly process most of the reads
  which have both ends close together in the BAM file while avoiding the use
  of excessive amounts of main memory to handle those pairs which are not
  close together.\pb

  \ifarxiv
  \begin{figure}
  \begin{center}
  \resizebox{.5\textwidth}{!}{\includegraphics{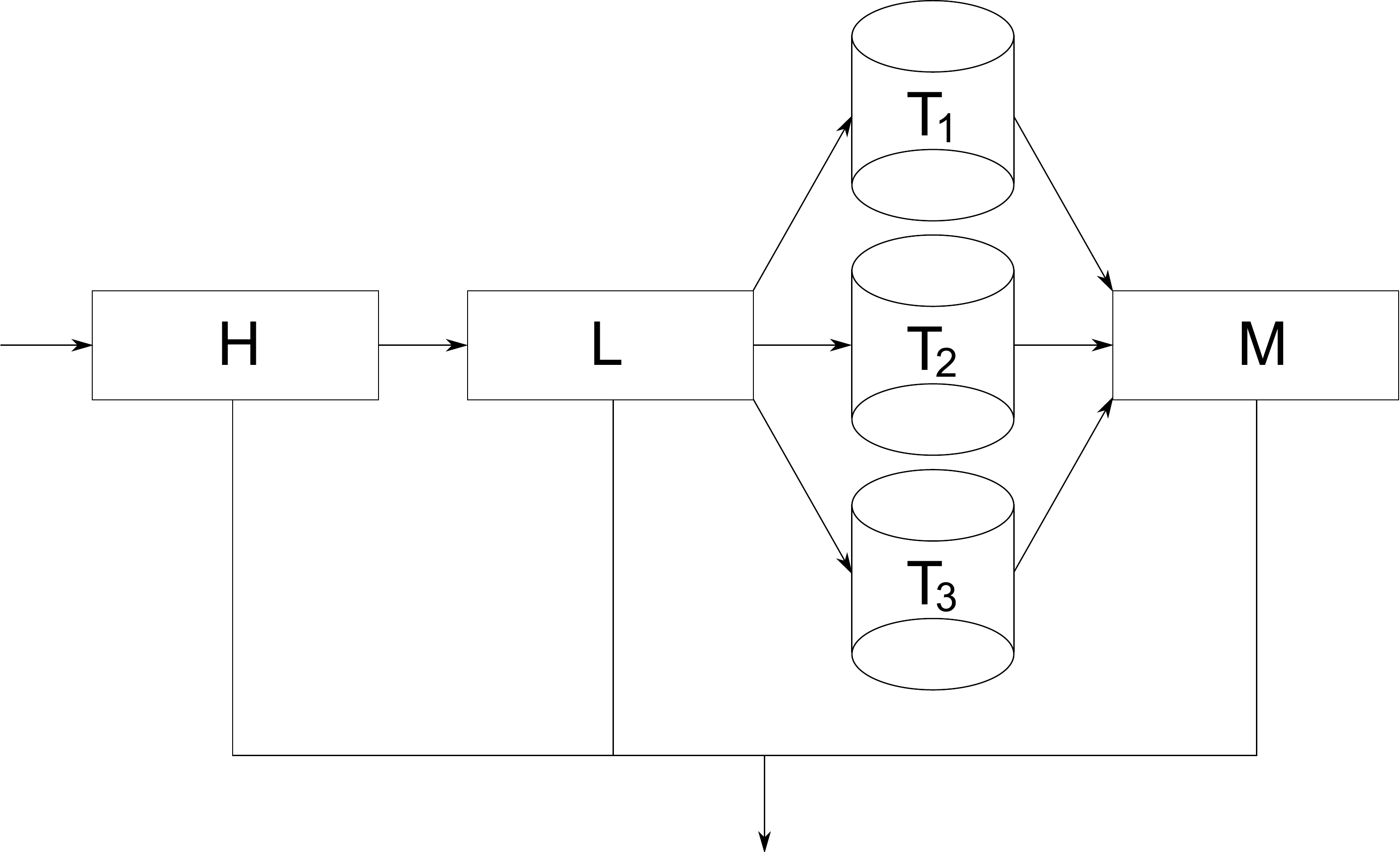}}
  \caption{
  Data Flow during Collation - The collation process uses several layers of data structures
  for handling alignments. This includes the hash table $H$ (see Figure
  $2$), the overflow list $L$ (see Figure $3$), a set of temporary files
  $T_i$ and a merged list $M$ produced from the $T_i$.
  }
  \label{figure1}
  \end{center}
  \end{figure}
  \fi

  To avoid the overhead resulting from the allocation of a small block of
  memory for each single alignment, we implement the hash table $H$ and the
  list $L$ in the following way. The hash table $H$ is implemented as a
  fixed size character array $R$ which we use as a ring buffer, an array $P$ of
  integers and a B-tree $B$. $P$ is the actual hash table storing pointers into $R$, 
  $R$ is used to store alignments as uncompressed BAM entries
  and $B$ contains the starting positions of all alignments currently stored in $R$.
  A pointer $r$ which is initially set to $0$ marks the current position in $R$. When
  a name $q$ is to be searched in $H$, then we first compute the hash value $h$ of
  the name and check whether position $P[h]$ in $R$ designates the start of
  an alignment and the stored alignment has the name $q$. An alignment with
  hash value $h$ can be erased from $H$ by first removing $P[h]$ from $B$ and then setting $P[h]$ 
  to a special value marking a free position in $P$. To insert a new
  alignment with hash value $h$ into $H$ we first need to make sure there is sufficient space.
  If $P[h]$ is used, then the currently stored alignment for $h$ needs to be
  moved to $L$ and erased from $H$. Then we possibly need to remove more
  alignments from $H$ until the difference between the current insert
  pointer $r$ and the next higher value in $B$ (considered in a circular way
  as $R$ is a ring buffer) contains sufficient space to store the new
  alignment. As soon as sufficient space is available, we can copy the
  alignment data to position $r$ in $R$, insert $r$ into $B$, store $P[h]=r$
  and advance $r$ by the length of the alignment data we have just stored.
  Figure $2$ visualises the components of the hash table $H$. We store the
  list $L$ as a byte array. The alignment data is filled in at the front end
  of the array. The pointers to the alignment starting positions in the byte
  array are filled in from the back of the array. The list runs full if we
  are no longer able to add the next alignment in the same way as the ones
  already stored. Figure $3$ shows a list $L$ containing $5$ alignments
  $A_0,A_1,\ldots,A_4$. A full list can be sorted by keeping the alignment
  data in place and reordering the pointers at the end of the byte array.
  Storing $H$ and $L$ in this way requires a very small amount of memory allocation
  and freeing operations for handling large sets of alignments.
  \ifarxiv
  \begin{figure}
  \begin{center}
  \resizebox{.5\textwidth}{!}{\includegraphics{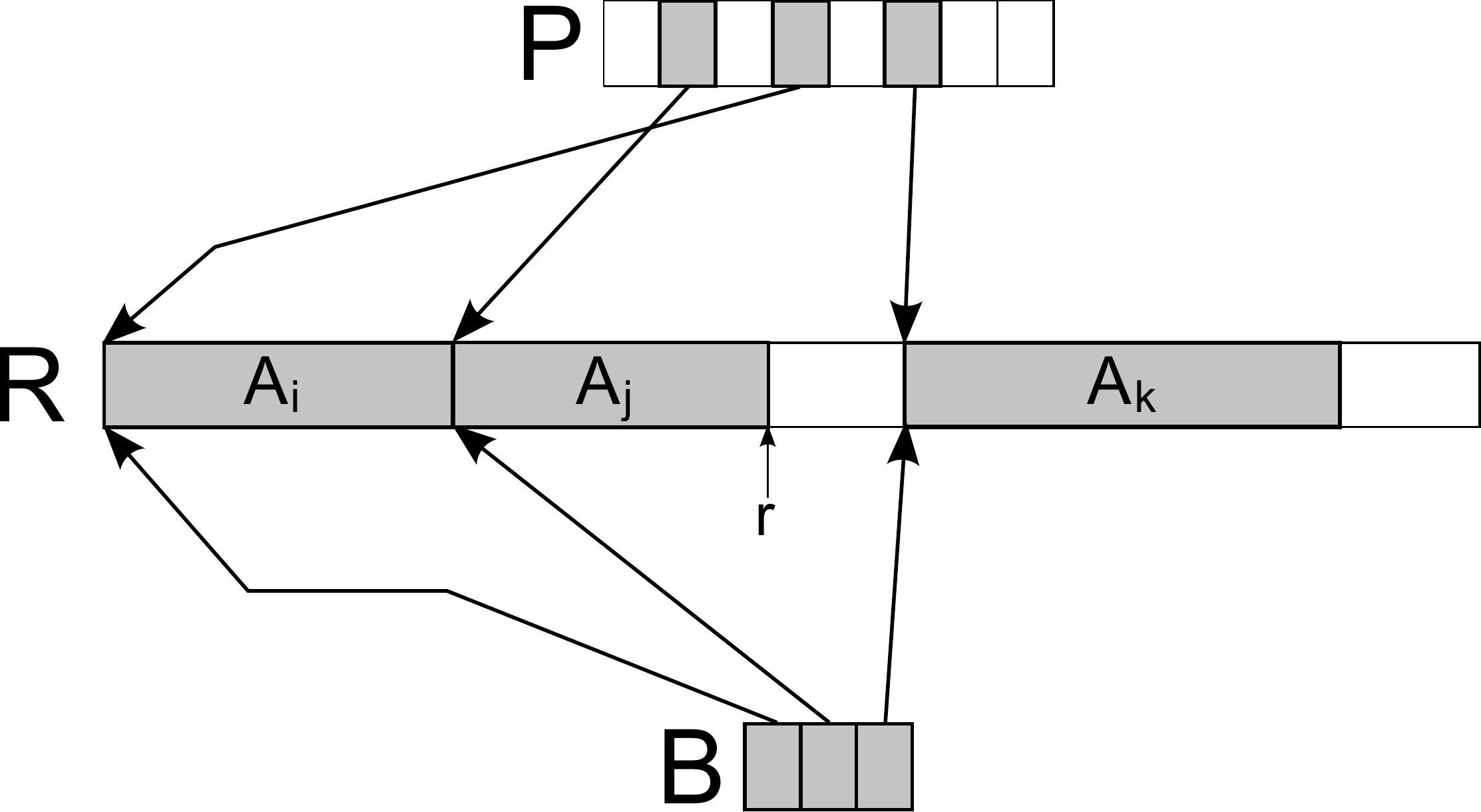}}
  \caption{
  Collation Hash Table $H$ - The hash table $H$ used for collation is composed of three components. The
  ring buffer $R$ stores alignment data. In the picture it contains three
  alignments $A_i, A_j$ and $A_k$. The insert pointer $r$ is situated just
  after the alignment $A_j$. The hash table $P$ stores pointers into $R$,
  where the position of the respective pointers is given by a hash value
  computed from the name of the stored alignment. The B-tree $B$ stores the
  starting positions of alignments in $R$ in sorted order.
  }
  \label{figure2}
  \end{center}
  \end{figure}
  \fi
  \ifarxiv
  \begin{figure}
  \begin{center}
  \resizebox{.5\textwidth}{!}{\includegraphics{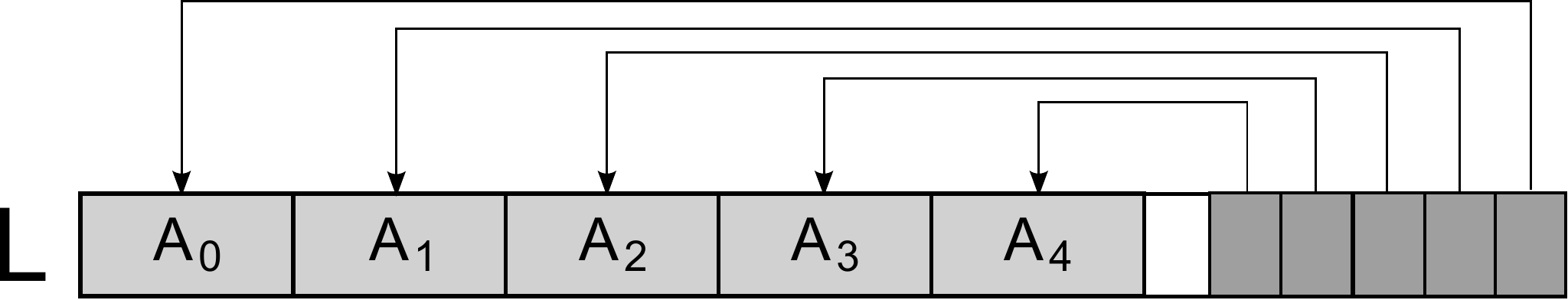}}
  \caption{
  Overflow List $L$ - The overflow list $L$ is implemented as a byte array. Alignments are
  inserted from the start of the array. In the picture $A_0, A_1, \ldots
  A_4$ are contained. Pointers to the respective starting positions are
  inserted from the end of the byte array.
  }
  \label{figure3}
  \end{center}
  \end{figure}
  \fi

  \section*{Results and Discussion}

  \subsection*{Using \texttt{libmaus}/\texttt{biobambam}: a short introduction}

  In the following we will describe how to obtain, compile and use the
  \texttt{libmaus}-API for collating reads extracted from BAM files by name
  and how to use the tools \texttt{bamtofastq} and
  \texttt{bammarkduplicates}.

  \subsection*{Installation}

  The installation of \texttt{libmaus} and \texttt{biobambam} is as fairly
  straight-forward process on a recent Linux system. 
  Both are based on the standard GNU
  autoconf/automake based build system. The latest source tarballs are
  available at \texttt{https://github.com/gt1/libmaus/tags} and
  \texttt{https://github.com/gt1/biobambam/tags} respectively. After unpacking
  the tarballs the packages can be compiled using
  \begin{alltt}
  autoreconf -i -f
  ./configure --prefix=\(\texttt{\$}\){HOME}/libmaus && make install
  \end{alltt}
  for libmaus and subsequently
  \begin{alltt}
  autoreconf -i -f
  ./configure --with-libmaus=\(\texttt{\$}\){HOME}/libmaus --prefix=\(\texttt{\$}\){HOME}/biobambam && make install
  \end{alltt}
  for the biobambam front-end tools. Using these commands the tools
  \texttt{bamtofastq} and \texttt{bammarkduplicates} will be installed in
  the directory \texttt{\${HOME}/biobambam/bin}.\pb
  The installation on recent versions of Ubuntu Linux (11.10 and newer) is 
  particularly easy, as both packages can be installed from LaunchPad as binaries:
  \begin{alltt}
  sudo add-apt-repository ppa:gt1/staden-io-lib-current
  sudo add-apt-repository ppa:gt1/libmaus
  sudo add-apt-repository ppa:gt1/biobambam
  sudo apt-get install libmaus-dev biobambam
  \end{alltt}
  This will place the new tools in the \texttt{/usr/bin} directory. The
  LaunchPad version comes with support for converting CRAM files to FastQ
  via the Staden package's \texttt{io\_lib}.\pb
  \texttt{libmaus} can be compiled on MacOS X in way very similar to the one shown for
  Linux. On non recent versions of the MacOS X development environment this will require
  the \texttt{boost} libaries (see \cite{boost}).
  \pb

  \subsection*{Compiling programs using \texttt{libmaus}}
  The compiler and linker flags necessary for using \texttt{libmaus} can be
  obtained using the \texttt{pkg-config} tool. If \texttt{libmaus} is not
  installed in a system directory via LaunchPad, then \texttt{pkg-config}
  needs to be informed of it's location via the \texttt{PKG\_CONFIG\_PATH}
  environment variable. An example for the \texttt{bash} shell is
  \begin{alltt}
  export PKG_CONFIG_PATH=\(\texttt{\$}\){HOME}/libmaus/lib/pkgconfig:\(\texttt{\$}\)PKG_CONFIG_PATH 
  \end{alltt}
  The compilation flags can then be obtained using
  \begin{alltt}
  pkg-config --cflags --libs libmaus
  \end{alltt}
  A sample program can thus be compiled using
  \begin{alltt}
  c++ source.cpp -o binary `pkg-config --cflags --libs libmaus`
  \end{alltt}
  \subsection*{Including read name collating BAM input in libmaus in source code}
  For including the name collating BAM input functionality of libmaus
  in other C++ source code, the respective definitions need to be made
  available using
  \begin{verbatim}
#include <libmaus/bambam/CircularHashCollatingBamDecoder.hpp>	
#include <libmaus/util/TempFileRemovalContainer.hpp>	
using namespace libmaus::bambam;
using namespace libmaus::util;
using namespace std;
typedef BamCircularHashCollatingBamDecoder collator_type;
typedef collator_type::alignment_ptr_type alignment_ptr_type;
  \end{verbatim}
  The first two lines include header files from \texttt{libmaus}. The next five 
  lines simplify notation in the following. The collating input class can
  then be instantiated using
  \begin{verbatim}
collator_type C(cin,"tmpfile");
  \end{verbatim}
to read from the standard input channel \texttt{cin}. The second argument
specifies the name of the file used to write alignments out to disk when the
list $L$ described above overflows. The temporary file can be removed after
all alignments have been extracted from the input stream. For the sake of
convenience this can also be done automatically using
  \begin{verbatim}
TempFileRemovalContainer::addTempFile("tmpfile");
\end{verbatim}
After the instantiation of the collator object pairs can be extracted using
  \begin{verbatim}
pair<alignment_ptr_type,alignment_ptr_type> P;
while ( C.tryPair(P) )
 if ( P.first && P.second )
 {
  /* process pair */
  cout << "Found pair with name " << P.first->getName() << endl;
 }
\end{verbatim}
The function \texttt{tryPair} of the collator class tries to extract
pairs from the input BAM file. It returns \texttt{true} if any
data could be extracted. The pair \texttt{P} will contain two pointers to
alignments if this extraction was succesful. In case there are single or
orphan reads in the input one of the two pointers may be a null pointer (an
orphan read is a read end such that the other end is missing from the file).
A list with accessor functions for alignments with their respective return
types is shown in Table 1. Header information like the length and name of
reference sequences can be obtained by calling methods of the header object
in the collation class.
\begin{verbatim}
BamHeader const & header = C.getHeader();
\end{verbatim}
Some methods and return types of the \texttt{BamHeader} class can be found
in Table 2.

\ifarxiv
\begin{table}
\begin{center}
      \begin{tabular}{|l|l|l|}
        \hline \multicolumn{3}{|c|}{Alignment accessor functions}\\ \hline
        \em{Name of function} & \em{Return type} & \em{Description} \\ \hline
        getName() & string & alignment name \\ \hline
        getLReadName() & integer & length of read name \\ \hline
        getRefID() & integer & id of reference sequence this end was mapped to \\ \hline
        getPos() & integer & position on reference sequence this end was mapped to \\ \hline
        getNextRefID() & integer & id of reference sequence other end was mapped to \\ \hline
        getNextPos() & integer & position on reference sequence other end was mapped to \\ \hline
        getFlags() & integer & alignment flags \\ \hline
	isPaired() & bool & true if read was paired in sequencing \\ \hline
	isProper() & bool & true if template is mapped as a proper pair \\ \hline
	isMapped() & bool & true if this end is mapped \\ \hline
	isMateMapped() & bool & true if other end is mapped \\ \hline
	isReverse() & bool & true if this end is mapped to the reverse strand \\ \hline
	isMateReverse() & bool & true if other end is mapped to the reverse strand \\ \hline
	isRead1() & bool & true if this end is the first read of the pair \\ \hline
	isRead2() & bool & true if this end is the second read of the pair \\ \hline
	isSecondary() & bool & true if this alignment is not the primary one \\ \hline
	isQCFail() & bool & true if alignment has failed quality control \\ \hline
	isDup() & bool & true if alignment is duplicate of another \\ \hline
        getLseq() & integer & length of query sequence \\ \hline
	getRead() & string & query sequence \\ \hline
	getReadRC() & string & reverse complement of query sequence \\ \hline
	getQual() & string & quality string \\ \hline
	getQualRC() & string & reverse quality string \\ \hline
        getMapQ() & integer & mapping quality for this end \\ \hline
        getNCigar() & integer & number of cigar operations \\ \hline
        getCigarFieldOpAsChar(i) & character & i'th cigar operator as character \\ \hline
        getCigarFieldLength(i) & integer & i'th cigar operation length \\ \hline
        getTlen() & integer & infered template length \\ \hline
        getAuxAsString("tagname") & string & content of auxiliary field with id tagname \\ \hline
        formatFastQ() & string & alignment converted to a FastQ entry \\ \hline
      \end{tabular}
    \caption{Accessor methods of the alignment class in \texttt{libmaus}}
    \label{table1}
\end{center}
\end{table}

\begin{table}
\begin{center}
      \begin{tabular}{|l|l|l|}
        \hline \multicolumn{3}{|c|}{BAM header accessor functions}\\ \hline
        \em{Name of function} & \em{Return type} & \em{Description} \\ \hline
        getRefIDName(i) & string & name of i'th reference sequence \\ \hline
        getRefIDLength(i) & integer & length of i'th reference sequence \\ \hline
        getNumRef() & integer & number of reference sequences \\ \hline
        getVersion() & string & BAM format version number \\ \hline
        getSortOrder() & string & sort order of the BAM file \\ \hline
      \end{tabular}

  \caption{Accessor methods of the BAM header class in \texttt{libmaus}}
\end{center}
\end{table}
\fi

  \subsection*{A sample program for converting BAM to FastQ}

The following code is a complete program for converting an input BAM file to
FastQ while keeping only complete pairs. 
\begin{verbatim}
#include <libmaus/bambam/CircularHashCollatingBamDecoder.hpp>	
#include <libmaus/util/TempFileRemovalContainer.hpp>

using namespace libmaus::bambam;
using namespace libmaus::util;
using namespace std;

int main()
{
 typedef BamCircularHashCollatingBamDecoder collator_type;
 typedef collator_type::alignment_ptr_type alignment_ptr_type;

 /* remove temporary file at program exit */
 string const tmpfilename = "tmpfile";
 TempFileRemovalContainer::addTempFile(tmpfilename);

 /* set up collator object */	
 collator_type C(cin,tmpfilename);
 pair<alignment_ptr_type,alignment_ptr_type> P;

 /* read alignments */	
 while ( C.tryPair(P) )
  /* if we have a pair, then print both ends as FastQ */
  if ( P.first && P.second )
  {
   cout << P.first->formatFastq();
   cout << P.second->formatFastq();
  }
}
\end{verbatim}
Note that the source code of the \texttt{bamtofastq} program in \texttt{biobambam} is
somewhat more complicated because it offers more options (different
input formats like SAM and CRAM, handling of single and orphan reads, etc.)
and introduces a few small syntatic nuances (for instance a reusable buffer
for the conversion of alignments to FastQ) to increase performance further.
The interested reader is refered to the respective source code (cf.~\cite{biobambam}).
  \subsubsection*{Performance Comparison of \texttt{bamtofastq} and Picard}
  All benchmarks for this papers were run on a PC using an
  Intel Core i$7$-$2700$K processor running at a frequency of $3.5$ GHz and equipped with $16$GB of memory.
  While this is a quad core CPU, the benchmarks were run using only $1$ thread per job.
  Temporary files were stored on a fast solid state type drive (SSD).
  The machine was running version $13.04$ (Raring Ringtail) of Ubuntu Linux.
  libmaus and biobambam were compiled using the version of the C++ compiler
  delivered with the operating system (GCC version $4.7.3$).
  We compare against version $1.91$ of the Picard suite. We have downloaded
  the binary distribution of Picard from SourceForge and run the programs 
  using Oracle's Java (Java SE 7u21).
  For evaluating the performance of our approach for converting BAM to FastQ in 
  comparison with Picard we have used the following data sets, each stored
  in a single BAM file:
  \begin{enumerate}
  \item 
	The low depth human data set HG00096 (see \cite{HG96}) from the 1000 genomes project
	(cf.~\cite{pmid20981092}) with a depth median value of $4$. We have used the BAM file as provided by the
	project, the size of the file is $15$GB. 
	The median of the distance between the two ends of a template in the file is $5$
	(due to some outliers the weighted average is $96220$).
	Both programs were able to handle the data set with a heap size of about $256$MB.
	The runtime was 
		$305$s ($5$m$5$s) for \texttt{bamtofastq}
	and 
		$1182$s ($19$m$42$s) for Picard's SamToFastq component, 
	i.e.~\texttt{bamtofastq}
	was faster by a factor of $3.87$.
        None of the two programs gained significantly by using more main memory.
	%
	%%%
	%
  \item 
	The high depth human data set ERP001231 (see \cite{ERP001231}) with a median depth value of 45.
	This data was downloaded as FastQ and mapped to the human reference 
        \cite{GRCh37} using the SMALT aligner \cite{smalt}. 
        The resulting BAM file was sorted by coordinate using SAMtools.
	The median of the distance between the two ends of a template in the file is $107$
	but due to some very pronounced peeks and outliers in the distribution the weighted average is $1.4\cdot 10^7$.
	The size of the resulting sorted BAM file is $95$GB. 
	\texttt{bamtofastq} was able to handle the file in $2661$s ($44$m$21$s) seconds using its default memory setting of
	$256$MB of heap space. Using more main memory resulted in only
	slightly better perfomance, the runtime dropped to $42$m$28$s for $816$MB of heap
	space and did not decrease further for $1451$MB.
	Picard ran out of memory on the file when provided with $5$GB and less of main memory.
	It took $30270$s ($8$h$24$m) given $6$GB of main memory.
	By increasing the amount of memory to $10$GB the runtime was reduced to $9870$s ($2$h$44$m).
	Due to the modest drop in run time from $9$GB ($2$h$47$m) to $10$GB we did
	not measure the run time for more than $10$GB.
	These numbers give \texttt{bamtofastq} a runtime which is between $3.78$ and
	$11.88$ times faster than Picard's MarkDuplicates.
	The high runtime of Picard using $6$GB of main memory suggests that
	that the employed memory management gets very inefficient when the
	program uses most of the provided memory. For $5$GB it fails after a
	run time of more than 7 hours.
	% Check for higher
	%A further increase (we tried up to $16$GB) did not yield a
	%significantly lower runtime. This gives \texttt{bamtofastq} a runtime
	%which is between $6$ and $7$ times faster than Picard depending on memory settings.
	%
	%%%
	%
  \item
	The high depth E.~coli data set from the study SRP017681 (see \cite{SRP017681}) with a median depth value of $879$.
	This data was also downloaded as FastQ and mapped to the respective reference genome (see \cite{K12}) using SMALT.
        The resulting BAM file was sorted by coordinate using SAMtools. The sorted BAM file has a size of $44$GB.
	The median of the distance between the two ends of a template in the file is $158240$, the weighted average is $1.7\cdot 10^7$.
	\texttt{bamtofastq} is able to handle this file in $2689$s ($44$m$49$s) using $256$MB of main memory.
        In this memory setting a large amount of reads need to be handled by
	resorting to temporary files on secondary storage because of the high depth of the input data.
	%Because of the high depth of the data a large amount of reads need to be handled by resorting
	%to temporary files on secondary storage. 
	Due to this effect the runtime decreases to $1720$s ($28$m$40$s) when we let the program use $1.45$G
	of memory by increasing the size of the hash table $H$. 
	Picard fails with an out of memory type error when given $8$GB of main memory. 
	Using $9$GB of memory it processes the file in $4$h$7$m. Increasing the
	main memory given to $10$G and $11$G decreases the run time to $2$h$25$m and $2$h$6$m respecively. 
	Thus \texttt{bamtofastq} is depending on the memory settings between $2.8$ and $8.6$ times faster 
	than Picard while using significantly less memory.
	%
	%%%
	%
  \end{enumerate}
 Picard uses Java's \texttt{HashMap} class and keeps each end in this
 hash table until the other end of the read is found in the file. This
 explains its high memory requirements. The performance is also low due to 
 frequent object allocation and implicit deallocation (garbage collection)
 processes, in particular when the memory used is close to the memory given.
 \texttt{bamtofastq} can handle all the given files easily with its default 
 small memory foot print. In particular it does not require the user to
 adjust the input parameters to process any of the files.
 \subsubsection*{Marking Duplicate Alignments}
 Large sets of sequenced reads often contain reads or read pairs which are
 not unique, i.e.~such reads and read pairs which map to the same
 coordinates on a given reference genome. This may happen for several
 reasons including artefacts of library preparation (e.g.~duplication by
 PCR), sequencing artefacts (e.g. optical duplicates) or just by chance as
 the selection of sequenced molecules is usually a random process.
 For some data sets to number of duplicate reads can be very high.
 In the E.~coli data set SRP017681 mentioned above (see \cite{SRP017681}) for instance
 more than 90\% of the reads are duplicates. The presence of duplicates can
 significantly influence downstream analysis. Thus the detection and marking
 of duplicates is an important step in the analysis of sequenced data.
 The Picard tool suite contains a program for marking duplicates in BAM
 files. We will not describe the algorithmic procedure employed in detail
 here but only provide a rough description of how it works. 
 First the program constructs a list $L_P$ of aligned pairs and a list of
 aligned single ended reads and orphans $L_S$. Both lists are sorted by
 coordinates, where the sorting of the list $L_P$ is lexicographic in the
 coordinates of the two ends (i.e.~the pairs are first sorted by the
 leftmost mapping end and then those which have the same leftmost mapping
 position are sorted by the mapping coordinate of the other end).
 In this sorted order it is very simple to partition the lists $L_P$ and
 $L_S$ into subsets of read pairs and single reads respectively which map to
 the same coordinates. In each such subset a single element with the highest score
 computed from the base qualities of the reads is selected as representant
 and the other elements are considered and marked as duplicates. In addition
 the current code also considers single ended reads and orphans as duplicates 
 if they map to the same coordinate as one end of a mapped pair.
 The read name collation approach we present in this paper can be used as a
 building block for the construction of the list $L_P$. Based on this we
 have implemented our own version of a duplication marking tool
 \texttt{bammarkduplicates}, which in its first release mimics the behaviour
 of Picard's MarkDuplicates tools. In addition to the read name collation
 machinery it uses a structure very similar to the list $L$ above (see also
 Figure $3$) for sorting fragments of the lists $L_P$ and $L_S$ before those
 fragments are written to secondary memory. This gives
 \texttt{bammarkduplicates} a very stable and predictable memory usage
 profile. This is different from Picard's MarkDuplicates tool, which uses a
 large amount of main memory for some data sets featuring high coverage in
 some regions or as a whole and thus is harder to handle in automated
 sequencing pipelines, as it sometimes requires manual intervention due to
 out of memory type errors.\pb

 We have evaluated the performance of \texttt{bammarkduplicates} in
 comparison to Picard's MarkDuplicates for the same BAM files as we have
 used above for the BAM to FastQ evaluation.
 \begin{enumerate}
  \item As can be expected the low depth data set HG00096 can efficiently be
    handled by both programs using $256$MB of heap space. None of the two
    programs benefit from being allowed to use more memory.
    \texttt{bammarkduplicates} handles the file in $36$m$49$s and
    MarkDuplicate in $42$m$42$s. Thus on this data set
    \texttt{bammarkduplicates} is faster by a factor of $1.16$.
  \item 
	\texttt{bammarkduplicates} processes the file
	obtained by mapping the reads from ERP001231 as described above in
	$5$h$28$m using $281$MB of main memory. It does not benefit from
	using more memory.
	Picard is not able to
	handle the file using $512$MB of RAM. With $768$MB it runs for
	$5$h$37$m. The run time for $2$GB and $3$GB is about the same at
	$5$h$30$m, thus we assume that further increasing the amount of
	given RAM will not result in better performance. The run time on
	this data set is thus about the same for both programs, while
	\texttt{bammarkduplicates} uses less memory.
  \item
	The BAM file obtained by mapping the reads from the study SRP017681
	as stated above is handled by \texttt{bammarkduplicates} in time
	$2$h$45$m using $416$MB of main memory. The run time gradually
	decreases to $2$h$20$m when the main memory provided is increased to
	$2.56$GB. Picard's MarkDuplicates tool is not capable of handling
	the file given $3$GB of memory, it aborts with an out of memory
	type error. Given $4$GB it processes the file
	in time $2$h$29$m. Further increasing the main memory threshold to
	$10$GB decreases its run time to $2$h$23$m. Thus the run time of the
	two tools again is similar with a tendency of \texttt{bamtofastq}
	being slightly faster for an equivalent amount of given main memory.
	\texttt{bammarkduplicates} is capable of handling the file using
	significantly less memory than Picard.
 \end{enumerate}

%%%%%%%%%%%%%%%%%%%%%%
\section*{Conclusions}
 In this paper we have presented efficient algorithms and data structures
 for name collated BAM file input. We have provided an implementation of
 these in libmaus, an open source programming library for C++. As part of
 the biobambam package we have developed two tools \texttt{bamtofastq}
 and \texttt{bammarkduplicates} based on the API. These are often faster than
 their counter parts in Picard and use significantly less main memory.
  
%%%%%%%%%%%%%%%%%%%%%%%%%%%%%%%%
\section*{Competing interests}
  The authors declare that they have no competing interests.

\section*{Authors contributions}
  GT wrote the code, ran the tests and benchmarks and wrote the paper. 
  SL contributed to the testing and provided patches for making
  the bammarkduplicates tool more compatible with Picard's MarkDuplicates module.

%%%%%%%%%%%%%%%%%%%%%%%%%%%
\section*{Acknowledgements}
  \ifthenelse{\boolean{publ}}{\small}{}
  GT and SL are supported by the Wellcome Trust.

\section*{Availability and requirements}
  \ifthenelse{\boolean{publ}}{\small}{}

  \textbf{Project name:} biobambam/libmaus\\
  \textbf{Operating systems:} Linux and MacOS X\\
  \textbf{Programming language:} C++\\
  \textbf{Other requirements:} none\\
  \textbf{License:} GPL3\\
  \textbf{Any restrictions to use by non-academics:} none\\
 
%%  Bmc_article.bst  will be used to                       %%
%%  create a .BBL file for submission, which includes      %%
%%  XML structured for BMC.                                %%

{\ifthenelse{\boolean{publ}}{\footnotesize}{\small}
 \bibliographystyle{bmc_article}  % Style BST file
  \bibliography{biobambam} }     % Bibliography file (usually '*.bib' ) 

%%%%%%%%%%%

\ifthenelse{\boolean{publ}}{\end{multicols}}{}

%%%%%%%%%%%%%%%%%%%%%%%%%%%%%%%%%%%
%%                               %%
%% Figures                       %%
%%                               %%
%% NB: this is for captions and  %%
%% Titles. All graphics must be  %%
%% submitted separately and NOT  %%
%% included in the Tex document  %%
%%                               %%
%%%%%%%%%%%%%%%%%%%%%%%%%%%%%%%%%%%

%%
%% Do not use \listoffigures as most will included as separate files

\ifarxiv
\else
\section*{Figures}
  \subsection*{Figure 1 - Data Flow during Collation}
  The collation process uses several layers of data structures
  for handling alignments. This includes the hash table $H$ (see Figure
  $2$), the overflow list $L$ (see Figure $3$), a set of temporary files
  $T_i$ and a merged list $M$ produced from the $T_i$.

  \subsection*{Figure 2 - Collation Hash Table $H$}
  The hash table $H$ used for collation is composed of three components. The
  ring buffer $R$ stores alignment data. In the picture it contains three
  alignments $A_i, A_j$ and $A_k$. The insert pointer $r$ is situated just
  after the alignment $A_j$. The hash table $P$ stores pointers into $R$,
  where the position of the respective pointers is given by a hash value
  computed from the name of the stored alignment. The B-tree $B$ stores the
  starting positions of alignments in $R$ in sorted order.

  \subsection*{Figure 3 - Overflow List $L$}
  The overflow list $L$ is implemented as a byte array. Alignments are
  inserted from the start of the array. In the picture $A_0, A_1, \ldots
  A_4$ are contained. Pointers to the respective starting positions are
  inserted from the end of the byte array.

%%%%%%%%%%%%%%%%%%%%%%%%%%%%%%%%%%%
%%                               %%
%% Tables                        %%
%%                               %%
%%%%%%%%%%%%%%%%%%%%%%%%%%%%%%%%%%%

%% Use of \listoftables is discouraged.
%%
\section*{Tables}
  \subsection*{Table 1 - Alignment accessor functions}
    Accessor methods of the alignment class in \texttt{libmaus}
     \par \mbox{}
    \par
    \mbox{
      \begin{tabular}{|l|l|l|}
        \hline \multicolumn{3}{|c|}{Alignment accessor functions}\\ \hline
        \em{Name of function} & \em{Return type} & \em{Description} \\ \hline
        getName() & string & alignment name \\ \hline
        getLReadName() & integer & length of read name \\ \hline
        getRefID() & integer & id of reference sequence this end was mapped to \\ \hline
        getPos() & integer & position on reference sequence this end was mapped to \\ \hline
        getNextRefID() & integer & id of reference sequence other end was mapped to \\ \hline
        getNextPos() & integer & position on reference sequence other end was mapped to \\ \hline
        getFlags() & integer & alignment flags \\ \hline
	isPaired() & bool & true if read was paired in sequencing \\ \hline
	isProper() & bool & true if template is mapped as a proper pair \\ \hline
	isMapped() & bool & true if this end is mapped \\ \hline
	isMateMapped() & bool & true if other end is mapped \\ \hline
	isReverse() & bool & true if this end is mapped to the reverse strand \\ \hline
	isMateReverse() & bool & true if other end is mapped to the reverse strand \\ \hline
	isRead1() & bool & true if this end is the first read of the pair \\ \hline
	isRead2() & bool & true if this end is the second read of the pair \\ \hline
	isSecondary() & bool & true if this alignment is not the primary one \\ \hline
	isQCFail() & bool & true if alignment has failed quality control \\ \hline
	isDup() & bool & true if alignment is duplicate of another \\ \hline
        getLseq() & integer & length of query sequence \\ \hline
	getRead() & string & query sequence \\ \hline
	getReadRC() & string & reverse complement of query sequence \\ \hline
	getQual() & string & quality string \\ \hline
	getQualRC() & string & reverse quality string \\ \hline
        getMapQ() & integer & mapping quality for this end \\ \hline
        getNCigar() & integer & number of cigar operations \\ \hline
        getCigarFieldOpAsChar(i) & character & i'th cigar operator as character \\ \hline
        getCigarFieldLength(i) & integer & i'th cigar operation length \\ \hline
        getTlen() & integer & infered template length \\ \hline
        getAuxAsString("tagname") & string & content of auxiliary field with id tagname \\ \hline
        formatFastQ() & string & alignment converted to a FastQ entry \\ \hline
      \end{tabular}
      }

  \subsection*{Table 2 - BAM header accessor functions}
    Accessor methods of the BAM header class in \texttt{libmaus}
     \par \mbox{}
    \par
    \mbox{
      \begin{tabular}{|l|l|l|}
        \hline \multicolumn{3}{|c|}{BAM header accessor functions}\\ \hline
        \em{Name of function} & \em{Return type} & \em{Description} \\ \hline
        getRefIDName(i) & string & name of i'th reference sequence \\ \hline
        getRefIDLength(i) & integer & length of i'th reference sequence \\ \hline
        getNumRef() & integer & number of reference sequences \\ \hline
        getVersion() & string & BAM format version number \\ \hline
        getSortOrder() & string & sort order of the BAM file \\ \hline
      \end{tabular}
      }
\fi

\end{bmcformat}
\end{document}